\documentclass[aps,prl,twocolumn,superscriptaddress]{revtex4}

\usepackage{graphicx}
\usepackage{amsmath} 
\usepackage{amssymb}
\usepackage{array}
\usepackage{epstopdf}

\begin{document}
\def\kfeas{KFe$_2$As$_2$}
\def\ca{CaFe$_2$As$_2$}
\def\sr{SrFe$_2$As$_2$}
\def\ba{BaFe$_2$As$_2$}
\def\tc{$T_{c}$}
\def\pc{$P_c$}
\def\pM{$P_M$}
\def\hc2{$H_{c2}$}
\def\bak{(Ba,K)Fe$_2$As$_2$}
\def\ie{{\it i.e.}}
\def\ie{{\it et al.}}


\title{High-temperature superconductivity stabilized by electron-hole interband coupling\\ in collapsed tetragonal phase of KFe$_2$As$_2$ under high pressure}


\author{Yasuyuki~Nakajima}
\author{Renxiong~Wang}
\author{Tristin~Metz}
\author{Xiangfeng~Wang}
\author{Limin~Wang}
 \affiliation{Center for Nanophysics and Advanced Materials, Department of Physics, University of Maryland, College Park, Maryland 20742}
\author{Hyunchae~Cynn}
\author{Samuel~T.~Wier}
\author{Jason~R.~Jeffries}
\affiliation{Lawrence Livermore National Laboratory, 7000 East Avenue Livermore, CA 94550}
\author{Johnpierre~Paglione}
 \affiliation{Center for Nanophysics and Advanced Materials, Department of Physics, University of Maryland, College Park, Maryland 20742}

\date{\today}

\begin{abstract}
We report a high-pressure study of simultaneous low-temperature electrical resistivity and Hall effect measurements on high quality single-crystalline KFe$_2$As$_2$ using designer diamond anvil cell techniques with applied pressures up to 33~GPa. 
In the low pressure regime, we show that the superconducting transition temperature $T_{c}$ finds a maximum onset value of 7~K near 2~GPa, in contrast to previous reports that find a minimum $T_{c}$ and reversal of pressure dependence at this pressure. 
Upon applying higher pressures, this $T_{c}$ is diminished until a sudden drastic enhancement occurs coincident with a first-order structural phase transition into a collapsed tetragonal phase. The appearance of a distinct superconducting phase above 13~GPa is also accompanied by a sudden reversal of dominant charge carrier sign, from hole- to electron-like, which agrees with our band structure calculations predicting the emergence of an electron pocket and diminishment of hole pockets upon Fermi surface reconstruction. 
Our results suggest the high-temperature superconducting phase in KFe$_2$As$_2$ is substantially enhanced by the presence of nested electron and hole pockets, providing the key ingredient of high-{\tc} superconductivity in iron pnictide superconductors.

\end{abstract}


\maketitle


Superconductivity in iron-based compounds has introduced a new paradigm in our understanding of unconventional pairing mechanisms in which the repulsive interaction between different Fermi surfaces play a major role \cite{mazin08}. In the iron superconductors, a wide versatility is found in the symmetry of the superconducting (SC) gap function, including sign-reversed full gap ($s_{\pm}$) \cite{hashi09a,teras09a,luo09,nakay09} and symmetry-imposed ($d$) \cite{reid12} or accidental nodal states (nodal $s_{\pm}$) \cite{okaza12,watan14}. These paring symmetries have indeed been considered theoretically and experimentally \cite{pagli10,chubu12}, and can undergo a transition from one to another by chemical substitution or pressure \cite{tanat10a,tafti13,tafti15,watan14}. Capturing universal traits in these symmetries is widely thought to give us the key to understanding high-temperature superconductivity in these fascinating materials.

Located at the end of phase diagram in hole-doped {\bak} \cite{rotte08}, the stoichiometric intermetallic compound {\kfeas} is a promising platform for exploring the evolution of rich pairing symmetries in iron-pnictide superconductors. While in {\bak}, the gap symmetry is believed to be of the fully-gapped $s$-wave type \cite{hashi09,nakay09}, in {\kfeas} both symmetry-imposed \cite{reid12} and accidental nodal \cite{okaza12,watan14} gap functions have been reported, which suggests a transition or crossover of SC gap function with chemical doping. In addition to chemical manipulation, recent pressure studies on {\kfeas} have proposed a possible symmetry change from $d$- to $s_{\pm}$-wave state to explain the sudden reversal of {\tc} pressure dependence at {\pc} $\sim$ 2~GPa \cite{tafti13,tafti15}. 

Here we report a high pressure study of transport and structural properties of {\kfeas} up to 33 GPa, far beyond previous work. We find two striking features: first, an initial enhancement of {\tc} with pressure reaches up to 7 K around {\pc}, opposite to that observed in previous pressure work using hydrostatic pressure cells \cite{budko12,tafti13,teras14,taufo14,tafti15}. Second, upon applying higher pressure, we reveal another SC phase with maximum {\tc} of $\sim$11~K at 15~GPa that is more than double that of {\tc} = 4~K at ambient pressure, which emerges together with a sign change of the dominant charge carrier type as well as a drastic change in the transport properties. The switch of carrier type from hole- to electron-like is associated with a strong first-order structural collapse of the tetragonal unit cell to a collapsed tetragonal (cT) phase above $\sim$13~GPa, as confirmed by X-ray measurements and supported by electronic band structure calculations. The simultaneous observation of carrier switch and enhancement of {\tc} possibly suggests that interband coupling between hole and electron pockets is the key ingredient of superconductivity in high-{\tc} iron-pnictide superconductors.

High quality single crystals of {\kfeas} were obtained by KAs flux method, and placed in contact with the electrical microprobes of an eight-probe designer diamond anvil cell \cite{Weir00} 
configured to allow combinations of both longitudinal and transverse four-wire resistance measurements. Pressures were determined from the shift of the ruby fluorescence line \cite{vos91}.
Steatite powder was used as the pressure medium. Transport measurements $(I \parallel ab, H \parallel c)$ were performed in a dilution refrigerator. Structural studies were performed at Sector 16 BM-D (HPCAT) of the Advanced Photon Source using a microfocused (5$\times$12$\mu$m), 30-keV incident X-ray beam. The X-ray sample was cut with a razor blade and loaded into the sample chamber ($40\mu$m thickness, $130\mu$m diameter) of a DAC composed of 300-micron diamond anvils. Neon was used as the pressure-transmitting medium, and the pressure was calibrated with NaCl, a small amount of which was loaded into the chamber along with the DAC. Diffraction patterns were collected on a MAR345 image plate, and the data were integrated using the the program Fit2D \cite{hamme96}. Lattice parameters were extracted using JADE and EXPGUI/GSAS software packages \cite{toby01}.
Theoretical calculations were conducted using the WIEN2K \cite{schwa02} implementation of the full potential linearized augmented plane wave method in the local density approximation. The $k$-point mesh was taken to be 11$\times$11$\times$11.

\begin{figure}[tb]
\includegraphics[width=8.5cm]{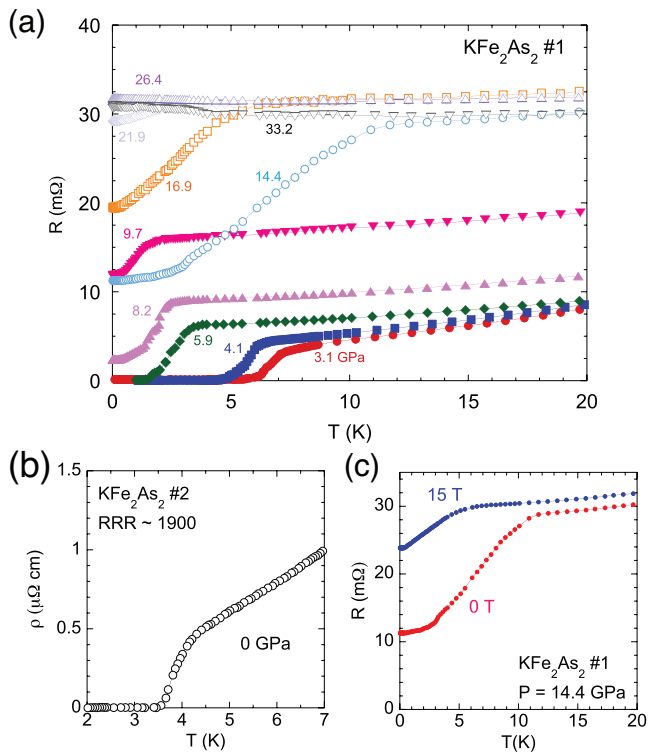}
\caption{Enhancement of {\tc} under pressure. a) Temperature dependence of resistance for {\kfeas} under pressure of sample \#1. b) Low temperature zoom of resistivity measured at ambient pressure (without DAC) of sample \#2 with RRR $\sim$ 1900. c) Suppression of superconducting transition at 14.4 GPa by applying magnetic field of 15 T.}
\end{figure}

As shown in Fig.~1, electrical resistivity measurements were performed on the same sample at different pressures using a four-wire configuration. (Resistance rather than resistivity values are reported due to a possible contamination from highly anisotropic $c$-axis conductance of the sample in the DAC; this is not the case in ambient pressure data, c.f. Fig.1b.)
At low pressures (\ie, $P<15$~GPa), these measurements reveal a surprising enhancement of {\tc} that is quite different from previous reports \cite{budko12,tafti13,teras14,taufo14,tafti15}. At ambient pressure, the zero-resistance state of {\kfeas} is observed at 3.6~K, followed by metallic behavior in the resistivity as shown in Fig.~1b. On applying pressure at our lowest value of 3.1 GPa, the transition temperature determined by zero resistance shows a sudden jump up to 6~K, followed by a a gradual reduction with increasing pressures up to 9.7 GPa that points to a complete termination of the superconducting phase near $\sim 15$~GPa. However, at 14.4 GPa we observe a sizable reduction of resistivity below 11~K that is strongly suppressed by applying magnetic field (see Fig.~1c), suggestive of a SC transition. Upon further pressure increase, the SC transition is gradually reduced, being completely suppressed to zero temperature near 26~GPa.


\begin{figure}[tb]
\includegraphics[width=8.5cm]{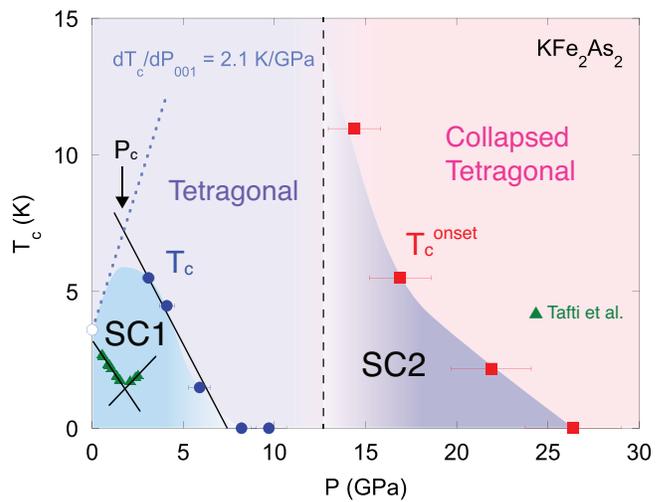}
\caption{$T-P$ phase diagram of {\kfeas}, with two distinct superconducting phases occupying different pressure regimes. The superconducting transition temperature $T_c$ of SC1 phase (solid circles) is determined from the zero resistance state, and $T_c^{onset}$ (filled squares) in the high-pressure SC2 phase determined from the onset of resistive transition for sample \#1  and \#2 (open symbols) in Figs.~1a and b, respectively. The critical pressure {\pc} is defined as the pressure where the sign reversal of {\tc} (triangles) in the low-pressure phase dependence has been observed previously \cite{tafti13}. The second superconducting phase SC2 appears above a structural collapse of the tetragonal unit cell at 13~GPa (see text). Blue dotted line is the pressure derivative of {\tc} obtained from the Ehrenfest relation (see text).}
\end{figure}

\begin{figure}[thb]
\includegraphics[width=8.5cm]{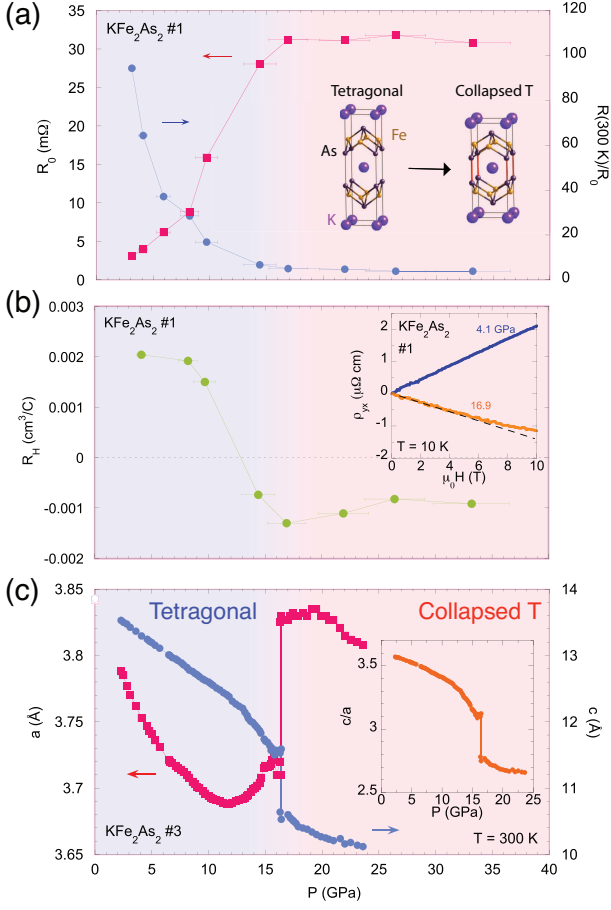}
\caption{(a) Pressure dependence of the residual resistance $R_0$ and the residual resistance ratio $R$(300~K)/$R_0$ for {\kfeas} (sample \#1). Inset shows the schematic structural change from tetragonal to collapsed tetragonal phase. (b) Hall coefficient obtained from the low field Hall resistivity at 10 K (12 K for 14.4 GPa) as a function of pressure. Inset: typical Hall resistivity data at 10K below and above the structural collapse. Dashed line indicates deviation from linear single-carrier expectation. (c) Room-temperature X-ray diffraction data, showing sudden decrease of the $c$-axis lattice parameter and increase of $a$ around 16 GPa, associated with a collapsed tetragonal transition. Open symbols at ambient pressure are obtained from Ref. \onlinecite{rozsa81}. Inset: pressure dependence of $c/a$.}
\end{figure}

We depict the pressure phase diagram of {\kfeas} with two SC phases in Fig.~2, defining {\tc} from the zero resistance (circles) and onset of the resistive transition (squares). 
For comparison, we also plot {\tc} from Tafti {\it et al.} \cite{tafti13}, showing the sudden reversal in pressure dependence of {\tc} previously reported. In the current work, we observe a peak in the low-pressure SC phase (SC1) that appears at the previously reported critical pressure {\pc}, rather than a minimum as observed in other studies \cite{budko12,tafti13,teras14,taufo14,tafti15}.
The enhancement (rather than suppression) of {\tc} in SC1 can be possibly explained by the reduction in hydrostatic pressure conditions produced by the steatite powder in our DAC experiment, as compared to the previous clamp-cell experiments. While the level of hydrostaticity at low pressures is generally considered good in this configuration, the sensitivity of {\tc} in {\kfeas} to uniaxial components may be susceptible to such differences and may explain the variation of {\tc} values reported at low pressures \cite{budko12,tafti13,teras14,taufo14,tafti15}.

We can estimate the slope of {\tc} as a function of uniaxial pressure by using the thermodynamic relation for the second-order transition, or the Ehrenfest relation \cite{barro99,hardy09},
\begin{equation}
	dT_c/dP_{001} = V_m\Delta\alpha_{001} T_c /\Delta C_p,
\end{equation}
where $V_m$ is the molar volume ($6.1\times$10$^{-5}$ m$^3$), $\Delta\alpha_{001}$ is the jump in the thermal expansion coefficient along $c$-axis at the phase transition, and $\Delta C_p$ is the jump in the heat capacity. Using the experimental data, $\Delta\alpha_{001} =$ 1.8 $\times$ 10$^-6$ K$^{-1}$ and $\Delta C_p/T_c =$ 54 mJ/mol K$^2$ obtained from Ref.~\onlinecite{burge13}, we extract a positive pressure derivative, $dT_c/dP_{001} =$ 2.1~K/GPa, drawn as a dashed line in Fig.~2. Surprisingly, the calculated $T_c(P)$ dependence crosses the linear extrapolation of our measured SC1 {\tc}data very close to {\pc}. The observation of a maximum at {\pc}, rather than a minimum, suggests that, if there is a SC order parameter symmetry change, then it must be first-order in nature. Alternatively, a crossover is occurring without symmetry change, or {\pc} is correlated with the recent photoemission study revealing the presence of a van Hove singularity \cite{fang14}.

\begin{figure}[tb]
\includegraphics[width=8.5cm]{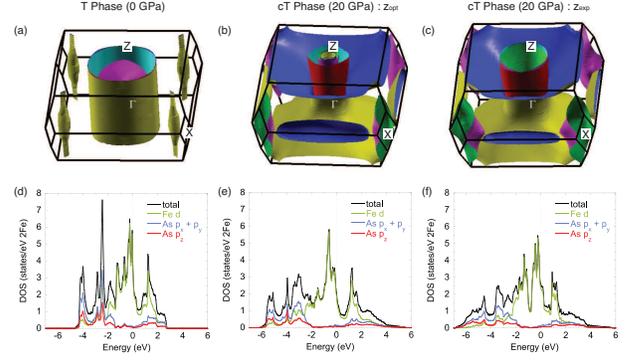}
\caption{Theoretical band calculations in {\kfeas}. Fermi surfaces in (a) tetragonal (T) phase (0 GPa), (b) collapsed tetragonal (cT) phase (20 GPa) obtained from the optimized arsenic height $z_{opt}=0.366$, and (c) from preliminarily experimental arsenic height $z_{exp}=0.386$. Density of states in (d) T phase and cT phase obtained from (e) $z_{opt}$ and (f) from $z_{exp}$.}
\end{figure}

Accompanied by the large enhancement of {\tc} in SC2, the charge transport in the normal state drastically changes around a critical pressure $P_{cT}$, as shown by the plateau in residual resistance $R_0$ shown in Fig.~3a. Note that the drastic increase of $R_0$ is not due to a reduction of cross section with pressure, but to the pressure response of $c$-axis component in the measured resistance. In contrast to the previous work reporting the gradual decrease of in-plane resistivity with pressure \cite{taufo14}, a three-fold increase in $R_0$ between 3 and 9 GPa is observed, inconsistent with the estimated increase of $R_0$ by 10\% from the change of geometric factor based on the measured variation of lattice parameters (Fig.~3c). 
Corresponding to the increase of $R_0$, the residual resistance ratio (RRR) $R$(300~K)$/R_0$ saturates at $\sim 4$ around $P_{cT}$, after showing a strong suppression with pressure. The strong suppression of RRR in SC2 suggests a change in the dominant scattering mechanism at low temperatures. To have a closer look at the abrupt change of transport properties in SC2, we show the measured Hall resistivity at 10 K in the inset of FIG.~3b. In SC1 ($P$ = 4.1 GPa), the Hall resistivity is linear in $H$ below 10 T and its sign is positive. However, in SC2 (P=16.9 GPa), the Hall resistivity is negative, strongly indicating a switch of dominant carriers from hole to electron, but the curvature evinced at higher fields suggests the persistence of remnant positive carriers. The dramatic switch is highlighted in the pressure dependence of Hall coefficient $R_H$ obtained from a linear fit to the Hall resistivity below 5 T (FIG.~3b). Intersecting $R_H=0$ at a pressure near 13~GPa, the positive Hall coefficient become negative and nearly constant in SC2. Together with the huge enhancement of {\tc} in SC2, the switch of dominant carrier is suggestive of a structural/electronic transition around $P_{cT}$.

Structural parameters under pressure reveal the prominent phase transition between the normal state in SC1 and SC2. At room temperature, we observe an abrupt enhancement of the lattice parameter $a$ and a sudden reduction of $c$ at 16~GPa without any crystallographic symmetry change, as extracted from X-ray diffraction data under pressure (FIG.~3c). Combined with no crystallographic symmetry change, the reduction of $c/a$ in the inset of Fig.~3c indicates the presence of a transition from tetragonal to cT phase at 16~GPa and 300~K, similar to the observation in other 122 system \cite{kreys08,goldm09,mitta11,saha12}. Although this collapsed transition induces the observed change in the transport properties, it does not cause the sizable enhancement of coupling between electrons and lattice, which could drive the observed high-{\tc} phase in SC2. To investigate the lattice properties, we can extract the bulk modulus $B_0$ at zero pressure and its pressure derivative $B_{0}^{\prime}$ from the pressure dependence of unit cell volume (not shown) by using the third order Birch-Murnaghan equation of state \cite{birch47}, 
\begin{align}
	P = \frac{3B_0}{2}[(V/V_0)^{-\frac{7}{3}}-(V/V_0)^{-\frac{5}{3}}]\nonumber\\
	\times\left[ 1+\frac{3(B_0^{\prime}-4)}{4}((V/V_0)^{-\frac{2}{3}}-1 )\right ],
\end{align}
where $V_0$ is the unit cell volume at ambient pressure. Fixing $B_0^{\prime}$ = 4 in both the tetragonal and cT phases, we obtain parameters of $B_0$ = 40.1 GPa for the tetragonal phase, and $B_0$ = 50.7 GPa for the cT phase. The increase of $B_0$ is rather small, compared with the large enhancement of $B_0$ observed in other iron pnictides undergoing a cT transition \cite{mitta11}. It is likely that the slight increase of $B_0$ in the cT phase with the higher $T_c$ excludes the simple phonon-mediated weak-coupling superconductivity in SC2 as well as SC1.

Electronic structure calculations suggest that the dramatic change of transport properties observed in the cT phase originates from a reconstruction of electronic structure. The schematic band structures of {\kfeas} in the tetragonal (0 GPa) and cT (20 GPa) phases obtained from our calculations are shown in Fig.~4. We use the experimental structural parameters for the tetragonal and cT phases, comparing the theoretically optimized arsenic height $z_{opt}=0.366$ in the cT phase (Fig.~4(b)) to a preliminary measurement of the experimental value $z_{exp}=0.386$ in the cT phase (Fig.~4(c)). These give qualitatively similar band structures, with notable enhancement of electron cylindrical pockets centered around the $\Gamma$ point. In the tetragonal phase, the Fermi surfaces consist of large hole bands around the $\Gamma$ point and a small hole pocket around the $X$ point in the Brillouin zone. In contrast, in the cT phase the hole bands around $\Gamma$ point shrink, becoming more three dimensional in the band. The hole pocket around the $X$ point vanishes, and instead a cylindrical electron band appears. Note that the hole bands completely disappear around the $\Gamma$ point in the collapsed phase observed in other 122 systems \cite{kasah11,gofry14}. The appearance of electron bands induced by the collapsed transition support the switch of dominant carrier type observed in our experiment.

Besides the dominant carrier change, the appearance of an electron band gives new insight into the possible SC pairing scenario of {\kfeas} in SC2, namely, that associated with the interband coupling between hole and electron pockets. Provided the same pairing mechanism in SC1 and SC2, we can naively attribute the higher {\tc} in SC2 to the enhancement of the density of states (DOS). However, as shown in FIG.~4c, the calculated DOS at the Fermi energy diminishes in the collapsed phase. Rather, this reduction of DOS in SC2 implies an enhancement of pairing interaction to explain the enhancement of {\tc}. Superconducting pairing in iron pnictides with higher {\tc} values, such as in {\bak}, is believed to arise from interband coupling between hole and electron pockets, connected to each other by Fermi surface nesting \cite{mazin08,kurok08}. In this case, in the presence of both hole and electron bands in SC2 of {\kfeas}, the pairing interaction associated with the nesting between those bands could be enhanced, compared with the interaction in SC1 without electron pockets in the band structure. Such a pairing scenario is also supported by the fact there is no bulk SC phase in the collapsed phase of CaFe$_2$As$_2$ realized under pressure, where the disappearance of hole pockets at $\Gamma$ point has been confirmed \cite{kasah11,gofry14}. 

In summary, we have investigated the transport and structural properties of {\kfeas} under pressures up to 33 GPa, revealing the presence of two superconducting phases that appear distinct, but each showing strong enhancements in their transition temperature as a function of pressure. The first low-pressure phase exhibits a {\tc} enhancement that is possibly connected to a strong sensitivity to uniaxial pressure components, while the second, higher-$T_c$ phase abruptly appears upon collapse of the tetragonal structure at higher pressures. This strong enhancement of {\tc} is accompanied by a change in the dominant charge carrier sign induced by the structural collapse, and is explained by electronic structure modifications that yield coexistent electron and hole pockets with coupling that appears to favor high temperature superconductivity.

The authors acknowledge valuable discussions with R. L. Greene and L. Taillefer. Work at the University of Maryland was supported by AFOSR Grant FA9550-14-1-0332 and the Gordon and Betty Moore Foundation’s EPiQS Initiative through Grant GBMF4419.  Portions of this work were performed under LDRD (Tracking Code 14-ERD-041).  Lawrence Livermore National Laboratory is operated by Lawrence Livermore National Security, LLC, for the DOE, NNSA under Contract No. DE-AC52-07NA27344. HPCAT operations are supported by DOE-NNSA under Award No. DE-NA0001974 and DOE-BES under Award No. DE-FG02-99ER45775, with partial instrumentation funding by NSF. APS is supported by DOE-BES, under Contract No. DE-AC02-06CH11357. A portion of the beamtime was provided by the Carnegie DOE-Alliance Center (CDAC).





\bibliographystyle{apsrev}

\bibliography{KFe2As2}

\begin{thebibliography}{36}
\expandafter\ifx\csname natexlab\endcsname\relax\def\natexlab#1{#1}\fi
\expandafter\ifx\csname bibnamefont\endcsname\relax
  \def\bibnamefont#1{#1}\fi
\expandafter\ifx\csname bibfnamefont\endcsname\relax
  \def\bibfnamefont#1{#1}\fi
\expandafter\ifx\csname citenamefont\endcsname\relax
  \def\citenamefont#1{#1}\fi
\expandafter\ifx\csname url\endcsname\relax
  \def\url#1{\texttt{#1}}\fi
\expandafter\ifx\csname urlprefix\endcsname\relax\def\urlprefix{URL }\fi
\providecommand{\bibinfo}[2]{#2}
\providecommand{\eprint}[2][]{\url{#2}}

\bibitem[{\citenamefont{Mazin et~al.}(2008)\citenamefont{Mazin, Singh,
  Johannes, and Du}}]{mazin08}
\bibinfo{author}{\bibfnamefont{I.~I.} \bibnamefont{Mazin}},
  \bibinfo{author}{\bibfnamefont{D.~J.} \bibnamefont{Singh}},
  \bibinfo{author}{\bibfnamefont{M.~D.} \bibnamefont{Johannes}},
  \bibnamefont{and} \bibinfo{author}{\bibfnamefont{M.~H.} \bibnamefont{Du}},
  \bibinfo{journal}{Phys. Rev. Lett.} \textbf{\bibinfo{volume}{101}},
  \bibinfo{pages}{057003} (\bibinfo{year}{2008}).

\bibitem[{\citenamefont{Hashimoto
  et~al.}(2009{\natexlab{a}})\citenamefont{Hashimoto, Shibauchi, Kato, Ikada,
  Okazaki, Shishido, Ishikado, Kito, Iyo, Eisaki et~al.}}]{hashi09a}
\bibinfo{author}{\bibfnamefont{K.}~\bibnamefont{Hashimoto}},
  \bibinfo{author}{\bibfnamefont{T.}~\bibnamefont{Shibauchi}},
  \bibinfo{author}{\bibfnamefont{T.}~\bibnamefont{Kato}},
  \bibinfo{author}{\bibfnamefont{K.}~\bibnamefont{Ikada}},
  \bibinfo{author}{\bibfnamefont{R.}~\bibnamefont{Okazaki}},
  \bibinfo{author}{\bibfnamefont{H.}~\bibnamefont{Shishido}},
  \bibinfo{author}{\bibfnamefont{M.}~\bibnamefont{Ishikado}},
  \bibinfo{author}{\bibfnamefont{H.}~\bibnamefont{Kito}},
  \bibinfo{author}{\bibfnamefont{A.}~\bibnamefont{Iyo}},
  \bibinfo{author}{\bibfnamefont{H.}~\bibnamefont{Eisaki}},
  \bibnamefont{et~al.}, \bibinfo{journal}{Phys. Rev. Lett.}
  \textbf{\bibinfo{volume}{102}}, \bibinfo{pages}{017002}
  (\bibinfo{year}{2009}{\natexlab{a}}).

\bibitem[{\citenamefont{Terashima et~al.}(2009)\citenamefont{Terashima, Sekiba,
  Bowen, Nakayama, Kawahara, Sato, Richard, Xu, Li, Cao et~al.}}]{teras09a}
\bibinfo{author}{\bibfnamefont{K.}~\bibnamefont{Terashima}},
  \bibinfo{author}{\bibfnamefont{Y.}~\bibnamefont{Sekiba}},
  \bibinfo{author}{\bibfnamefont{J.~H.} \bibnamefont{Bowen}},
  \bibinfo{author}{\bibfnamefont{K.}~\bibnamefont{Nakayama}},
  \bibinfo{author}{\bibfnamefont{T.}~\bibnamefont{Kawahara}},
  \bibinfo{author}{\bibfnamefont{T.}~\bibnamefont{Sato}},
  \bibinfo{author}{\bibfnamefont{P.}~\bibnamefont{Richard}},
  \bibinfo{author}{\bibfnamefont{Y.-M.} \bibnamefont{Xu}},
  \bibinfo{author}{\bibfnamefont{L.~J.} \bibnamefont{Li}},
  \bibinfo{author}{\bibfnamefont{G.~H.} \bibnamefont{Cao}},
  \bibnamefont{et~al.}, \bibinfo{journal}{Proceedings of the National Academy
  of Sciences} \textbf{\bibinfo{volume}{106}}, \bibinfo{pages}{7330}
  (\bibinfo{year}{2009}).

\bibitem[{\citenamefont{Luo et~al.}(2009)\citenamefont{Luo, Tanatar, Reid,
  Shakeripour, Doiron-Leyraud, Ni, Bud'ko, Canfield, Luo, Wang et~al.}}]{luo09}
\bibinfo{author}{\bibfnamefont{X.~G.} \bibnamefont{Luo}},
  \bibinfo{author}{\bibfnamefont{M.~A.} \bibnamefont{Tanatar}},
  \bibinfo{author}{\bibfnamefont{J.-P.} \bibnamefont{Reid}},
  \bibinfo{author}{\bibfnamefont{H.}~\bibnamefont{Shakeripour}},
  \bibinfo{author}{\bibfnamefont{N.}~\bibnamefont{Doiron-Leyraud}},
  \bibinfo{author}{\bibfnamefont{N.}~\bibnamefont{Ni}},
  \bibinfo{author}{\bibfnamefont{S.~L.} \bibnamefont{Bud'ko}},
  \bibinfo{author}{\bibfnamefont{P.~C.} \bibnamefont{Canfield}},
  \bibinfo{author}{\bibfnamefont{H.}~\bibnamefont{Luo}},
  \bibinfo{author}{\bibfnamefont{Z.}~\bibnamefont{Wang}}, \bibnamefont{et~al.},
  \bibinfo{journal}{Phys. Rev. B} \textbf{\bibinfo{volume}{80}},
  \bibinfo{pages}{140503} (\bibinfo{year}{2009}).

\bibitem[{\citenamefont{Nakayama et~al.}(2009)\citenamefont{Nakayama, Sato,
  Richard, Xu, Sekiba, Souma, Chen, Luo, Wang, Ding et~al.}}]{nakay09}
\bibinfo{author}{\bibfnamefont{K.}~\bibnamefont{Nakayama}},
  \bibinfo{author}{\bibfnamefont{T.}~\bibnamefont{Sato}},
  \bibinfo{author}{\bibfnamefont{P.}~\bibnamefont{Richard}},
  \bibinfo{author}{\bibfnamefont{Y.-M.} \bibnamefont{Xu}},
  \bibinfo{author}{\bibfnamefont{Y.}~\bibnamefont{Sekiba}},
  \bibinfo{author}{\bibfnamefont{S.}~\bibnamefont{Souma}},
  \bibinfo{author}{\bibfnamefont{G.~F.} \bibnamefont{Chen}},
  \bibinfo{author}{\bibfnamefont{J.~L.} \bibnamefont{Luo}},
  \bibinfo{author}{\bibfnamefont{N.~L.} \bibnamefont{Wang}},
  \bibinfo{author}{\bibfnamefont{H.}~\bibnamefont{Ding}}, \bibnamefont{et~al.},
  \bibinfo{journal}{EPL (Europhysics Letters)} \textbf{\bibinfo{volume}{85}},
  \bibinfo{pages}{67002} (\bibinfo{year}{2009}).

\bibitem[{\citenamefont{Reid et~al.}(2012)\citenamefont{Reid, Tanatar,
  Juneau-Fecteau, Gordon, de~Cotret, Doiron-Leyraud, Saito, Fukazawa, Kohori,
  Kihou et~al.}}]{reid12}
\bibinfo{author}{\bibfnamefont{J.-P.} \bibnamefont{Reid}},
  \bibinfo{author}{\bibfnamefont{M.~A.} \bibnamefont{Tanatar}},
  \bibinfo{author}{\bibfnamefont{A.}~\bibnamefont{Juneau-Fecteau}},
  \bibinfo{author}{\bibfnamefont{R.~T.} \bibnamefont{Gordon}},
  \bibinfo{author}{\bibfnamefont{S.~R.} \bibnamefont{de~Cotret}},
  \bibinfo{author}{\bibfnamefont{N.}~\bibnamefont{Doiron-Leyraud}},
  \bibinfo{author}{\bibfnamefont{T.}~\bibnamefont{Saito}},
  \bibinfo{author}{\bibfnamefont{H.}~\bibnamefont{Fukazawa}},
  \bibinfo{author}{\bibfnamefont{Y.}~\bibnamefont{Kohori}},
  \bibinfo{author}{\bibfnamefont{K.}~\bibnamefont{Kihou}},
  \bibnamefont{et~al.}, \bibinfo{journal}{Phys. Rev. Lett.}
  \textbf{\bibinfo{volume}{109}}, \bibinfo{pages}{087001}
  (\bibinfo{year}{2012}).

\bibitem[{\citenamefont{Okazaki et~al.}(2012)\citenamefont{Okazaki, Ota,
  Kotani, Malaeb, Ishida, Shimojima, Kiss, Watanabe, Chen, Kihou
  et~al.}}]{okaza12}
\bibinfo{author}{\bibfnamefont{K.}~\bibnamefont{Okazaki}},
  \bibinfo{author}{\bibfnamefont{Y.}~\bibnamefont{Ota}},
  \bibinfo{author}{\bibfnamefont{Y.}~\bibnamefont{Kotani}},
  \bibinfo{author}{\bibfnamefont{W.}~\bibnamefont{Malaeb}},
  \bibinfo{author}{\bibfnamefont{Y.}~\bibnamefont{Ishida}},
  \bibinfo{author}{\bibfnamefont{T.}~\bibnamefont{Shimojima}},
  \bibinfo{author}{\bibfnamefont{T.}~\bibnamefont{Kiss}},
  \bibinfo{author}{\bibfnamefont{S.}~\bibnamefont{Watanabe}},
  \bibinfo{author}{\bibfnamefont{C.-T.} \bibnamefont{Chen}},
  \bibinfo{author}{\bibfnamefont{K.}~\bibnamefont{Kihou}},
  \bibnamefont{et~al.}, \bibinfo{journal}{Science}
  \textbf{\bibinfo{volume}{337}}, \bibinfo{pages}{1314} (\bibinfo{year}{2012}).

\bibitem[{\citenamefont{Watanabe et~al.}(2014)\citenamefont{Watanabe,
  Yamashita, Kawamoto, Kurata, Mizukami, Ohta, Kasahara, Yamashita, Saito,
  Fukazawa et~al.}}]{watan14}
\bibinfo{author}{\bibfnamefont{D.}~\bibnamefont{Watanabe}},
  \bibinfo{author}{\bibfnamefont{T.}~\bibnamefont{Yamashita}},
  \bibinfo{author}{\bibfnamefont{Y.}~\bibnamefont{Kawamoto}},
  \bibinfo{author}{\bibfnamefont{S.}~\bibnamefont{Kurata}},
  \bibinfo{author}{\bibfnamefont{Y.}~\bibnamefont{Mizukami}},
  \bibinfo{author}{\bibfnamefont{T.}~\bibnamefont{Ohta}},
  \bibinfo{author}{\bibfnamefont{S.}~\bibnamefont{Kasahara}},
  \bibinfo{author}{\bibfnamefont{M.}~\bibnamefont{Yamashita}},
  \bibinfo{author}{\bibfnamefont{T.}~\bibnamefont{Saito}},
  \bibinfo{author}{\bibfnamefont{H.}~\bibnamefont{Fukazawa}},
  \bibnamefont{et~al.}, \bibinfo{journal}{Phys. Rev. B}
  \textbf{\bibinfo{volume}{89}}, \bibinfo{pages}{115112}
  (\bibinfo{year}{2014}).

\bibitem[{\citenamefont{Paglione and Greene}(2010)}]{pagli10}
\bibinfo{author}{\bibfnamefont{J.}~\bibnamefont{Paglione}} \bibnamefont{and}
  \bibinfo{author}{\bibfnamefont{R.~L.} \bibnamefont{Greene}},
  \bibinfo{journal}{Nat Phys} \textbf{\bibinfo{volume}{6}},
  \bibinfo{pages}{645} (\bibinfo{year}{2010}).

\bibitem[{\citenamefont{Chubukov}(2012)}]{chubu12}
\bibinfo{author}{\bibfnamefont{A.}~\bibnamefont{Chubukov}},
  \bibinfo{journal}{Annual Review of Condensed Matter Physics}
  \textbf{\bibinfo{volume}{3}}, \bibinfo{pages}{57} (\bibinfo{year}{2012}).

\bibitem[{\citenamefont{Tanatar et~al.}(2010)\citenamefont{Tanatar, Reid,
  Shakeripour, Luo, Doiron-Leyraud, Ni, Bud'ko, Canfield, Prozorov, and
  Taillefer}}]{tanat10a}
\bibinfo{author}{\bibfnamefont{M.~A.} \bibnamefont{Tanatar}},
  \bibinfo{author}{\bibfnamefont{J.-P.} \bibnamefont{Reid}},
  \bibinfo{author}{\bibfnamefont{H.}~\bibnamefont{Shakeripour}},
  \bibinfo{author}{\bibfnamefont{X.~G.} \bibnamefont{Luo}},
  \bibinfo{author}{\bibfnamefont{N.}~\bibnamefont{Doiron-Leyraud}},
  \bibinfo{author}{\bibfnamefont{N.}~\bibnamefont{Ni}},
  \bibinfo{author}{\bibfnamefont{S.~L.} \bibnamefont{Bud'ko}},
  \bibinfo{author}{\bibfnamefont{P.~C.} \bibnamefont{Canfield}},
  \bibinfo{author}{\bibfnamefont{R.}~\bibnamefont{Prozorov}}, \bibnamefont{and}
  \bibinfo{author}{\bibfnamefont{L.}~\bibnamefont{Taillefer}},
  \bibinfo{journal}{Phys. Rev. Lett.} \textbf{\bibinfo{volume}{104}},
  \bibinfo{pages}{067002} (\bibinfo{year}{2010}).

\bibitem[{\citenamefont{Tafti et~al.}(2013)\citenamefont{Tafti, Juneau-Fecteau,
  Delage, Rene~de Cotret, Reid, Wang, Luo, Chen, Doiron-Leyraud, and
  Taillefer}}]{tafti13}
\bibinfo{author}{\bibfnamefont{F.~F.} \bibnamefont{Tafti}},
  \bibinfo{author}{\bibfnamefont{A.}~\bibnamefont{Juneau-Fecteau}},
  \bibinfo{author}{\bibfnamefont{M.-E.} \bibnamefont{Delage}},
  \bibinfo{author}{\bibfnamefont{S.}~\bibnamefont{Rene~de Cotret}},
  \bibinfo{author}{\bibfnamefont{J.-P.} \bibnamefont{Reid}},
  \bibinfo{author}{\bibfnamefont{A.~F.} \bibnamefont{Wang}},
  \bibinfo{author}{\bibfnamefont{X.-G.} \bibnamefont{Luo}},
  \bibinfo{author}{\bibfnamefont{X.~H.} \bibnamefont{Chen}},
  \bibinfo{author}{\bibfnamefont{N.}~\bibnamefont{Doiron-Leyraud}},
  \bibnamefont{and}
  \bibinfo{author}{\bibfnamefont{L.}~\bibnamefont{Taillefer}},
  \bibinfo{journal}{Nat Phys} \textbf{\bibinfo{volume}{9}},
  \bibinfo{pages}{349} (\bibinfo{year}{2013}).

\bibitem[{\citenamefont{Tafti et~al.}(2015)\citenamefont{Tafti, Ouellet,
  Juneau-Fecteau, Faucher, Lapointe-Major, Doiron-Leyraud, Wang, Luo, Chen, and
  Taillefer}}]{tafti15}
\bibinfo{author}{\bibfnamefont{F.~F.} \bibnamefont{Tafti}},
  \bibinfo{author}{\bibfnamefont{A.}~\bibnamefont{Ouellet}},
  \bibinfo{author}{\bibfnamefont{A.}~\bibnamefont{Juneau-Fecteau}},
  \bibinfo{author}{\bibfnamefont{S.}~\bibnamefont{Faucher}},
  \bibinfo{author}{\bibfnamefont{M.}~\bibnamefont{Lapointe-Major}},
  \bibinfo{author}{\bibfnamefont{N.}~\bibnamefont{Doiron-Leyraud}},
  \bibinfo{author}{\bibfnamefont{A.~F.} \bibnamefont{Wang}},
  \bibinfo{author}{\bibfnamefont{X.-G.} \bibnamefont{Luo}},
  \bibinfo{author}{\bibfnamefont{X.~H.} \bibnamefont{Chen}}, \bibnamefont{and}
  \bibinfo{author}{\bibfnamefont{L.}~\bibnamefont{Taillefer}},
  \bibinfo{journal}{Phys. Rev. B} \textbf{\bibinfo{volume}{91}},
  \bibinfo{pages}{054511} (\bibinfo{year}{2015}).

\bibitem[{\citenamefont{Rotter et~al.}(2008)\citenamefont{Rotter, Pangerl,
  Tegel, and Johrendt}}]{rotte08}
\bibinfo{author}{\bibfnamefont{M.}~\bibnamefont{Rotter}},
  \bibinfo{author}{\bibfnamefont{M.}~\bibnamefont{Pangerl}},
  \bibinfo{author}{\bibfnamefont{M.}~\bibnamefont{Tegel}}, \bibnamefont{and}
  \bibinfo{author}{\bibfnamefont{D.}~\bibnamefont{Johrendt}},
  \bibinfo{journal}{Angewandte Chemie International Edition}
  \textbf{\bibinfo{volume}{47}}, \bibinfo{pages}{7949} (\bibinfo{year}{2008}).

\bibitem[{\citenamefont{Hashimoto
  et~al.}(2009{\natexlab{b}})\citenamefont{Hashimoto, Shibauchi, Kasahara,
  Ikada, Tonegawa, Kato, Okazaki, van~der Beek, Konczykowski, Takeya
  et~al.}}]{hashi09}
\bibinfo{author}{\bibfnamefont{K.}~\bibnamefont{Hashimoto}},
  \bibinfo{author}{\bibfnamefont{T.}~\bibnamefont{Shibauchi}},
  \bibinfo{author}{\bibfnamefont{S.}~\bibnamefont{Kasahara}},
  \bibinfo{author}{\bibfnamefont{K.}~\bibnamefont{Ikada}},
  \bibinfo{author}{\bibfnamefont{S.}~\bibnamefont{Tonegawa}},
  \bibinfo{author}{\bibfnamefont{T.}~\bibnamefont{Kato}},
  \bibinfo{author}{\bibfnamefont{R.}~\bibnamefont{Okazaki}},
  \bibinfo{author}{\bibfnamefont{C.~J.} \bibnamefont{van~der Beek}},
  \bibinfo{author}{\bibfnamefont{M.}~\bibnamefont{Konczykowski}},
  \bibinfo{author}{\bibfnamefont{H.}~\bibnamefont{Takeya}},
  \bibnamefont{et~al.}, \bibinfo{journal}{Phys. Rev. Lett.}
  \textbf{\bibinfo{volume}{102}}, \bibinfo{pages}{207001}
  (\bibinfo{year}{2009}{\natexlab{b}}).

\bibitem[{\citenamefont{Bud'ko et~al.}(2012)\citenamefont{Bud'ko, Liu,
  Lograsso, and Canfield}}]{budko12}
\bibinfo{author}{\bibfnamefont{S.~L.} \bibnamefont{Bud'ko}},
  \bibinfo{author}{\bibfnamefont{Y.}~\bibnamefont{Liu}},
  \bibinfo{author}{\bibfnamefont{T.~A.} \bibnamefont{Lograsso}},
  \bibnamefont{and} \bibinfo{author}{\bibfnamefont{P.~C.}
  \bibnamefont{Canfield}}, \bibinfo{journal}{Phys. Rev. B}
  \textbf{\bibinfo{volume}{86}}, \bibinfo{pages}{224514}
  (\bibinfo{year}{2012}).

\bibitem[{\citenamefont{Terashima et~al.}(2014)\citenamefont{Terashima, Kihou,
  Sugii, Kikugawa, Matsumoto, Ishida, Lee, Iyo, Eisaki, and Uji}}]{teras14}
\bibinfo{author}{\bibfnamefont{T.}~\bibnamefont{Terashima}},
  \bibinfo{author}{\bibfnamefont{K.}~\bibnamefont{Kihou}},
  \bibinfo{author}{\bibfnamefont{K.}~\bibnamefont{Sugii}},
  \bibinfo{author}{\bibfnamefont{N.}~\bibnamefont{Kikugawa}},
  \bibinfo{author}{\bibfnamefont{T.}~\bibnamefont{Matsumoto}},
  \bibinfo{author}{\bibfnamefont{S.}~\bibnamefont{Ishida}},
  \bibinfo{author}{\bibfnamefont{C.-H.} \bibnamefont{Lee}},
  \bibinfo{author}{\bibfnamefont{A.}~\bibnamefont{Iyo}},
  \bibinfo{author}{\bibfnamefont{H.}~\bibnamefont{Eisaki}}, \bibnamefont{and}
  \bibinfo{author}{\bibfnamefont{S.}~\bibnamefont{Uji}},
  \bibinfo{journal}{Phys. Rev. B} \textbf{\bibinfo{volume}{89}},
  \bibinfo{pages}{134520} (\bibinfo{year}{2014}).

\bibitem[{\citenamefont{Taufour et~al.}(2014)\citenamefont{Taufour, Foroozani,
  Tanatar, Lim, Kaluarachchi, Kim, Liu, Lograsso, Kogan, Prozorov
  et~al.}}]{taufo14}
\bibinfo{author}{\bibfnamefont{V.}~\bibnamefont{Taufour}},
  \bibinfo{author}{\bibfnamefont{N.}~\bibnamefont{Foroozani}},
  \bibinfo{author}{\bibfnamefont{M.~A.} \bibnamefont{Tanatar}},
  \bibinfo{author}{\bibfnamefont{J.}~\bibnamefont{Lim}},
  \bibinfo{author}{\bibfnamefont{U.}~\bibnamefont{Kaluarachchi}},
  \bibinfo{author}{\bibfnamefont{S.~K.} \bibnamefont{Kim}},
  \bibinfo{author}{\bibfnamefont{Y.}~\bibnamefont{Liu}},
  \bibinfo{author}{\bibfnamefont{T.~A.} \bibnamefont{Lograsso}},
  \bibinfo{author}{\bibfnamefont{V.~G.} \bibnamefont{Kogan}},
  \bibinfo{author}{\bibfnamefont{R.}~\bibnamefont{Prozorov}},
  \bibnamefont{et~al.}, \bibinfo{journal}{Phys. Rev. B}
  \textbf{\bibinfo{volume}{89}}, \bibinfo{pages}{220509}
  (\bibinfo{year}{2014}).

\bibitem[{\citenamefont{Weir et~al.}(2000)\citenamefont{Weir, Akella,
  Aracne-Ruddle, Vohra, and Catledge}}]{Weir00}
\bibinfo{author}{\bibfnamefont{S.~T.} \bibnamefont{Weir}},
  \bibinfo{author}{\bibfnamefont{J.}~\bibnamefont{Akella}},
  \bibinfo{author}{\bibfnamefont{C.}~\bibnamefont{Aracne-Ruddle}},
  \bibinfo{author}{\bibfnamefont{Y.~K.} \bibnamefont{Vohra}}, \bibnamefont{and}
  \bibinfo{author}{\bibfnamefont{S.~A.} \bibnamefont{Catledge}},
  \bibinfo{journal}{Applied Physics Letters} \textbf{\bibinfo{volume}{77}},
  \bibinfo{pages}{3400} (\bibinfo{year}{2000}).

\bibitem[{\citenamefont{Vos and Schouten}(1991)}]{vos91}
\bibinfo{author}{\bibfnamefont{W.~L.} \bibnamefont{Vos}} \bibnamefont{and}
  \bibinfo{author}{\bibfnamefont{J.~A.} \bibnamefont{Schouten}},
  \bibinfo{journal}{Journal of Applied Physics} \textbf{\bibinfo{volume}{69}},
  \bibinfo{pages}{6744} (\bibinfo{year}{1991}).

\bibitem[{\citenamefont{Hammersley et~al.}(1996)\citenamefont{Hammersley,
  Svensson, Hanfland, Fitch, and Hausermann}}]{hamme96}
\bibinfo{author}{\bibfnamefont{A.~P.} \bibnamefont{Hammersley}},
  \bibinfo{author}{\bibfnamefont{S.~O.} \bibnamefont{Svensson}},
  \bibinfo{author}{\bibfnamefont{M.}~\bibnamefont{Hanfland}},
  \bibinfo{author}{\bibfnamefont{A.~N.} \bibnamefont{Fitch}}, \bibnamefont{and}
  \bibinfo{author}{\bibfnamefont{D.}~\bibnamefont{Hausermann}},
  \bibinfo{journal}{High Pressure Research} \textbf{\bibinfo{volume}{14}},
  \bibinfo{pages}{235} (\bibinfo{year}{1996}).

\bibitem[{\citenamefont{Toby}(2001)}]{toby01}
\bibinfo{author}{\bibfnamefont{B.~H.} \bibnamefont{Toby}},
  \bibinfo{journal}{Journal of Applied Crystallography}
  \textbf{\bibinfo{volume}{34}}, \bibinfo{pages}{210} (\bibinfo{year}{2001}).

\bibitem[{\citenamefont{Schwarz et~al.}(2002)\citenamefont{Schwarz, Blaha, and
  Madsen}}]{schwa02}
\bibinfo{author}{\bibfnamefont{K.}~\bibnamefont{Schwarz}},
  \bibinfo{author}{\bibfnamefont{P.}~\bibnamefont{Blaha}}, \bibnamefont{and}
  \bibinfo{author}{\bibfnamefont{G.~K.~H.} \bibnamefont{Madsen}},
  \bibinfo{journal}{Comput. Phys. Commun} \textbf{\bibinfo{volume}{147}},
  \bibinfo{pages}{71} (\bibinfo{year}{2002}).

\bibitem[{\citenamefont{Rozsa and Schuster}(1981)}]{rozsa81}
\bibinfo{author}{\bibfnamefont{S.}~\bibnamefont{Rozsa}} \bibnamefont{and}
  \bibinfo{author}{\bibfnamefont{H.~U.} \bibnamefont{Schuster}},
  \bibinfo{journal}{Z. Naturforsch. B} \textbf{\bibinfo{volume}{36}},
  \bibinfo{pages}{1668} (\bibinfo{year}{1981}).

\bibitem[{\citenamefont{Barron and White}(1999)}]{barro99}
\bibinfo{author}{\bibfnamefont{T.~H.~K.} \bibnamefont{Barron}}
  \bibnamefont{and} \bibinfo{author}{\bibfnamefont{G.~K.} \bibnamefont{White}},
  \emph{\bibinfo{title}{Heat Capacity and Thermal Expansion at Low
  Temperatures}} (\bibinfo{publisher}{Kluwer Academic/Plenum, New York},
  \bibinfo{year}{1999}).

\bibitem[{\citenamefont{Hardy et~al.}(2009)\citenamefont{Hardy, Adelmann, Wolf,
  v.~L\"ohneysen, and Meingast}}]{hardy09}
\bibinfo{author}{\bibfnamefont{F.}~\bibnamefont{Hardy}},
  \bibinfo{author}{\bibfnamefont{P.}~\bibnamefont{Adelmann}},
  \bibinfo{author}{\bibfnamefont{T.}~\bibnamefont{Wolf}},
  \bibinfo{author}{\bibfnamefont{H.}~\bibnamefont{v.~L\"ohneysen}},
  \bibnamefont{and} \bibinfo{author}{\bibfnamefont{C.}~\bibnamefont{Meingast}},
  \bibinfo{journal}{Phys. Rev. Lett.} \textbf{\bibinfo{volume}{102}},
  \bibinfo{pages}{187004} (\bibinfo{year}{2009}).

\bibitem[{\citenamefont{Burger et~al.}(2013)\citenamefont{Burger, Hardy, Aoki,
  B\"ohmer, Eder, Heid, Wolf, Schweiss, Fromknecht, Jackson et~al.}}]{burge13}
\bibinfo{author}{\bibfnamefont{P.}~\bibnamefont{Burger}},
  \bibinfo{author}{\bibfnamefont{F.}~\bibnamefont{Hardy}},
  \bibinfo{author}{\bibfnamefont{D.}~\bibnamefont{Aoki}},
  \bibinfo{author}{\bibfnamefont{A.~E.} \bibnamefont{B\"ohmer}},
  \bibinfo{author}{\bibfnamefont{R.}~\bibnamefont{Eder}},
  \bibinfo{author}{\bibfnamefont{R.}~\bibnamefont{Heid}},
  \bibinfo{author}{\bibfnamefont{T.}~\bibnamefont{Wolf}},
  \bibinfo{author}{\bibfnamefont{P.}~\bibnamefont{Schweiss}},
  \bibinfo{author}{\bibfnamefont{R.}~\bibnamefont{Fromknecht}},
  \bibinfo{author}{\bibfnamefont{M.~J.} \bibnamefont{Jackson}},
  \bibnamefont{et~al.}, \bibinfo{journal}{Phys. Rev. B}
  \textbf{\bibinfo{volume}{88}}, \bibinfo{pages}{014517}
  (\bibinfo{year}{2013}).

\bibitem[{\citenamefont{Fang et~al.}(2014)\citenamefont{Fang, Shi, Du, Richard,
  Yang, X.X.Wu, Zhang, Qian, Ding, Wang et~al.}}]{fang14}
\bibinfo{author}{\bibfnamefont{D.}~\bibnamefont{Fang}},
  \bibinfo{author}{\bibfnamefont{X.}~\bibnamefont{Shi}},
  \bibinfo{author}{\bibfnamefont{Z.}~\bibnamefont{Du}},
  \bibinfo{author}{\bibfnamefont{P.}~\bibnamefont{Richard}},
  \bibinfo{author}{\bibfnamefont{H.}~\bibnamefont{Yang}},
  \bibinfo{author}{\bibnamefont{X.X.Wu}},
  \bibinfo{author}{\bibfnamefont{P.}~\bibnamefont{Zhang}},
  \bibinfo{author}{\bibfnamefont{T.}~\bibnamefont{Qian}},
  \bibinfo{author}{\bibfnamefont{X.}~\bibnamefont{Ding}},
  \bibinfo{author}{\bibfnamefont{Z.}~\bibnamefont{Wang}}, \bibnamefont{et~al.},
  \bibinfo{journal}{arXiv:1412.0945}  (\bibinfo{year}{2014}).

\bibitem[{\citenamefont{Kreyssig et~al.}(2008)\citenamefont{Kreyssig, Green,
  Lee, Samolyuk, Zajdel, Lynn, Bud'ko, Torikachvili, Ni, Nandi
  et~al.}}]{kreys08}
\bibinfo{author}{\bibfnamefont{A.}~\bibnamefont{Kreyssig}},
  \bibinfo{author}{\bibfnamefont{M.~A.} \bibnamefont{Green}},
  \bibinfo{author}{\bibfnamefont{Y.}~\bibnamefont{Lee}},
  \bibinfo{author}{\bibfnamefont{G.~D.} \bibnamefont{Samolyuk}},
  \bibinfo{author}{\bibfnamefont{P.}~\bibnamefont{Zajdel}},
  \bibinfo{author}{\bibfnamefont{J.~W.} \bibnamefont{Lynn}},
  \bibinfo{author}{\bibfnamefont{S.~L.} \bibnamefont{Bud'ko}},
  \bibinfo{author}{\bibfnamefont{M.~S.} \bibnamefont{Torikachvili}},
  \bibinfo{author}{\bibfnamefont{N.}~\bibnamefont{Ni}},
  \bibinfo{author}{\bibfnamefont{S.}~\bibnamefont{Nandi}},
  \bibnamefont{et~al.}, \bibinfo{journal}{Phys. Rev. B}
  \textbf{\bibinfo{volume}{78}}, \bibinfo{pages}{184517}
  (\bibinfo{year}{2008}).

\bibitem[{\citenamefont{Goldman et~al.}(2009)\citenamefont{Goldman, Kreyssig,
  Proke\ifmmode~\check{s}\else \v{s}\fi{}, Pratt, Argyriou, Lynn, Nandi,
  Kimber, Chen, Lee et~al.}}]{goldm09}
\bibinfo{author}{\bibfnamefont{A.~I.} \bibnamefont{Goldman}},
  \bibinfo{author}{\bibfnamefont{A.}~\bibnamefont{Kreyssig}},
  \bibinfo{author}{\bibfnamefont{K.}~\bibnamefont{Proke\ifmmode~\check{s}\else
  \v{s}\fi{}}}, \bibinfo{author}{\bibfnamefont{D.~K.} \bibnamefont{Pratt}},
  \bibinfo{author}{\bibfnamefont{D.~N.} \bibnamefont{Argyriou}},
  \bibinfo{author}{\bibfnamefont{J.~W.} \bibnamefont{Lynn}},
  \bibinfo{author}{\bibfnamefont{S.}~\bibnamefont{Nandi}},
  \bibinfo{author}{\bibfnamefont{S.~A.~J.} \bibnamefont{Kimber}},
  \bibinfo{author}{\bibfnamefont{Y.}~\bibnamefont{Chen}},
  \bibinfo{author}{\bibfnamefont{Y.~B.} \bibnamefont{Lee}},
  \bibnamefont{et~al.}, \bibinfo{journal}{Phys. Rev. B}
  \textbf{\bibinfo{volume}{79}}, \bibinfo{pages}{024513}
  (\bibinfo{year}{2009}).

\bibitem[{\citenamefont{Mittal et~al.}(2011)\citenamefont{Mittal, Mishra,
  Chaplot, Ovsyannikov, Greenberg, Trots, Dubrovinsky, Su, Brueckel, Matsuishi
  et~al.}}]{mitta11}
\bibinfo{author}{\bibfnamefont{R.}~\bibnamefont{Mittal}},
  \bibinfo{author}{\bibfnamefont{S.~K.} \bibnamefont{Mishra}},
  \bibinfo{author}{\bibfnamefont{S.~L.} \bibnamefont{Chaplot}},
  \bibinfo{author}{\bibfnamefont{S.~V.} \bibnamefont{Ovsyannikov}},
  \bibinfo{author}{\bibfnamefont{E.}~\bibnamefont{Greenberg}},
  \bibinfo{author}{\bibfnamefont{D.~M.} \bibnamefont{Trots}},
  \bibinfo{author}{\bibfnamefont{L.}~\bibnamefont{Dubrovinsky}},
  \bibinfo{author}{\bibfnamefont{Y.}~\bibnamefont{Su}},
  \bibinfo{author}{\bibfnamefont{T.}~\bibnamefont{Brueckel}},
  \bibinfo{author}{\bibfnamefont{S.}~\bibnamefont{Matsuishi}},
  \bibnamefont{et~al.}, \bibinfo{journal}{Phys. Rev. B}
  \textbf{\bibinfo{volume}{83}}, \bibinfo{pages}{054503}
  (\bibinfo{year}{2011}).

\bibitem[{\citenamefont{Saha et~al.}(2012)\citenamefont{Saha, Butch, Drye,
  Magill, Ziemak, Kirshenbaum, Zavalij, Lynn, and Paglione}}]{saha12}
\bibinfo{author}{\bibfnamefont{S.~R.} \bibnamefont{Saha}},
  \bibinfo{author}{\bibfnamefont{N.~P.} \bibnamefont{Butch}},
  \bibinfo{author}{\bibfnamefont{T.}~\bibnamefont{Drye}},
  \bibinfo{author}{\bibfnamefont{J.}~\bibnamefont{Magill}},
  \bibinfo{author}{\bibfnamefont{S.}~\bibnamefont{Ziemak}},
  \bibinfo{author}{\bibfnamefont{K.}~\bibnamefont{Kirshenbaum}},
  \bibinfo{author}{\bibfnamefont{P.~Y.} \bibnamefont{Zavalij}},
  \bibinfo{author}{\bibfnamefont{J.~W.} \bibnamefont{Lynn}}, \bibnamefont{and}
  \bibinfo{author}{\bibfnamefont{J.}~\bibnamefont{Paglione}},
  \bibinfo{journal}{Phys. Rev. B} \textbf{\bibinfo{volume}{85}},
  \bibinfo{pages}{024525} (\bibinfo{year}{2012}).

\bibitem[{\citenamefont{Birch}(1947)}]{birch47}
\bibinfo{author}{\bibfnamefont{F.}~\bibnamefont{Birch}},
  \bibinfo{journal}{Phys. Rev.} \textbf{\bibinfo{volume}{71}},
  \bibinfo{pages}{809} (\bibinfo{year}{1947}).

\bibitem[{\citenamefont{Kasahara et~al.}(2011)\citenamefont{Kasahara,
  Shibauchi, Hashimoto, Nakai, Ikeda, Terashima, and Matsuda}}]{kasah11}
\bibinfo{author}{\bibfnamefont{S.}~\bibnamefont{Kasahara}},
  \bibinfo{author}{\bibfnamefont{T.}~\bibnamefont{Shibauchi}},
  \bibinfo{author}{\bibfnamefont{K.}~\bibnamefont{Hashimoto}},
  \bibinfo{author}{\bibfnamefont{Y.}~\bibnamefont{Nakai}},
  \bibinfo{author}{\bibfnamefont{H.}~\bibnamefont{Ikeda}},
  \bibinfo{author}{\bibfnamefont{T.}~\bibnamefont{Terashima}},
  \bibnamefont{and} \bibinfo{author}{\bibfnamefont{Y.}~\bibnamefont{Matsuda}},
  \bibinfo{journal}{Phys. Rev. B} \textbf{\bibinfo{volume}{83}},
  \bibinfo{pages}{060505} (\bibinfo{year}{2011}).

\bibitem[{\citenamefont{Gofryk et~al.}(2014)\citenamefont{Gofryk, Saparov,
  Durakiewicz, Chikina, Danzenb\"acher, Vyalikh, Graf, and Sefat}}]{gofry14}
\bibinfo{author}{\bibfnamefont{K.}~\bibnamefont{Gofryk}},
  \bibinfo{author}{\bibfnamefont{B.}~\bibnamefont{Saparov}},
  \bibinfo{author}{\bibfnamefont{T.}~\bibnamefont{Durakiewicz}},
  \bibinfo{author}{\bibfnamefont{A.}~\bibnamefont{Chikina}},
  \bibinfo{author}{\bibfnamefont{S.}~\bibnamefont{Danzenb\"acher}},
  \bibinfo{author}{\bibfnamefont{D.}~\bibnamefont{Vyalikh}},
  \bibinfo{author}{\bibfnamefont{M.}~\bibnamefont{Graf}}, \bibnamefont{and}
  \bibinfo{author}{\bibfnamefont{A.}~\bibnamefont{Sefat}},
  \bibinfo{journal}{Phys. Rev. Lett.} \textbf{\bibinfo{volume}{112}},
  \bibinfo{pages}{186401} (\bibinfo{year}{2014}).

\bibitem[{\citenamefont{Kuroki et~al.}(2008)\citenamefont{Kuroki, Onari, Arita,
  Usui, Tanaka, Kontani, and Aoki}}]{kurok08}
\bibinfo{author}{\bibfnamefont{K.}~\bibnamefont{Kuroki}},
  \bibinfo{author}{\bibfnamefont{S.}~\bibnamefont{Onari}},
  \bibinfo{author}{\bibfnamefont{R.}~\bibnamefont{Arita}},
  \bibinfo{author}{\bibfnamefont{H.}~\bibnamefont{Usui}},
  \bibinfo{author}{\bibfnamefont{Y.}~\bibnamefont{Tanaka}},
  \bibinfo{author}{\bibfnamefont{H.}~\bibnamefont{Kontani}}, \bibnamefont{and}
  \bibinfo{author}{\bibfnamefont{H.}~\bibnamefont{Aoki}},
  \bibinfo{journal}{Phys. Rev. Lett.} \textbf{\bibinfo{volume}{101}},
  \bibinfo{pages}{087004} (\bibinfo{year}{2008}).

\end{thebibliography}

\end{document}